\documentclass[]{pasj01}

\begin{document} 
\Received{}
\Accepted{}

\title{Suzaku Observations of the Outskirts of the Galaxy Cluster Abell 3395 including a Filament toward Abell 3391}

%
%

 \author{
   Yuuki \textsc{Sugawara}\altaffilmark{1},
   Motokazu \textsc{Takizawa}\altaffilmark{2},
   Madoka \textsc{Itahana}\altaffilmark{1},
   Hiroki \textsc{Akamatsu}\altaffilmark{3},
   Yutaka \textsc{Fujita}\altaffilmark{4},
   Takaya \textsc{Ohashi}\altaffilmark{5},
   Yoshitaka \textsc{Ishisaki}\altaffilmark{5},
   }
 \altaffiltext{1}{School of Science and Engineering, Yamagata University, 
                  Kojirakawa-machi 1-4-12, Yamagata 990-8560, Japan}
  \altaffiltext{2}{Department of Physics, Yamagata University, Kojirakawa-machi
                  1-4-12, Yamagata 990-8560, Japan}
 \email{takizawa@sci.kj.yamagata-u.ac.jp}
  \altaffiltext{3}{SRON Netherlands Institute for Space Research, Sorbonnelaan 2, 3584 CA Utrecht, The Netherlands}
  \altaffiltext{4}{Department of Earth and Space Science, Graduate School of Science, Osaka University, Toyonaka, Osaka 560-0043, Japan}
  \altaffiltext{5}{Department of Physics, Tokyo Metropolitan University, 1-1 Minami-Osawa, Hachioji, Tokyo 192-0397, Japan}


\maketitle

\begin{abstract}
The results of Suzaku observations of the outskirts of Abell 3395 including a large-scale structure filament toward Abell 3391
are presented.
We measured temperature and abundance distributions from the southern outskirt of Abell 3395 to the north 
at the virial radius, where a filament structure has been found in the former X-ray and Sunyaev-Zel'dovich effect observations 
between Abell 3391 and 3395.
The overall temperature structure is consistent with the universal profile proposed by \citet{Okab14} for relaxed
clusters except for the filament region.
A hint of the ICM heating is found between the two clusters, which might be due to the interaction of them 
in the early phase of a cluster merger.
Although we obtained relatively low metal abundance of $Z=0.169^{+0.164+0.009+0.018 }_{-0.150-0.004-0.015 }$ solar, 
where the first, second, and third errors are statistical, cosmic X-ray background systematic, and non X-ray background systematic,
respectively, at the virial radius in the filament, our results are
still consistent with the former results of other clusters ($Z \sim 0.3$ solar) within errors. Therefore, our results are also
consistent with the early enrichment scenario.
We estimated Compton $y$ parameters only from X-ray results in the region between Abell 3391 and 3395 assuming a simple geometry. 
They are smaller than the previous SZ results with Planck satellite. 
The difference could be attributed to a more elaborate geometry such as a filament inclined to the line-of-sight direction,
or underestimation of the X-ray temperature because of the unresolved multi-temperature structures or undetected hot X-ray
emission of the shock heated gas.
\end{abstract}

\section{Introduction}
In the standard scenario of structure formation in the cold dark matter universe, 
clusters of galaxies form through mergers and absorptions of smaller groups, which occur more frequently
along cosmological large scale structure filaments. Main baryonic components in clusters of galaxies are in the
form of diffuse hot plasma called intracluster medium (ICM). It is well-known that ICM is polluted with 
heavy elements such as Fe, O, Si etc, 
which were originally synthesized in stars of galaxies, 
released to the interstellar space by supernovae explosions and mass losses of giant stars, 
and then transported to the intergalactic space by some mechanisms
(see \citet{Wern08} for a review).

Two likely scenarios are often considered for transporting heavy elements from galaxies
to the intergalactic space. 
One is due to ram pressure stripping 
\citep{Gunn72,Fuji99,Quil00,Doma06}.
It is obvious that this mechanism is more effective where the ambient gas density is higher and the relative velocity of a galaxy
to the ICM is larger. 
In other words, it is more effective in the central part of clusters and
after cluster galaxies assembled, which suggests relatively lower metal abundance in the cluster outskirts.
The other is a so-called galactic wind scenario, where a series of supernova explosions 
put energy into the interstellar matter and blow it outward \citep{DeYo78, Kepf06}. 
A similar situation is also expected owing to various activities of active galactic nuclei (AGN) 
\citep{Bens03,Cen06,Fabj10}.
Because these galactic activities are more prominent at $z \sim 2$ than the present universe 
\citep{Mada96,Fran01,Ueda03,Bran05},
heavy elements are expected to be transported into the intergalactic space before cluster formation in the galactic wind scenario.
In addition, this is more effective where the ambient gas pressure of galaxies is relatively low. In other words, 
it is more effective in the cluster outskirts and/or before cluster galaxies assembled.
As a result, relatively high metal abundance is expected in the cluster outskirts compared with the ram pressure stripping scenario.
Therefore, metal abundance in the outer part of clusters provides us with crucial information about metal transport processes 
from galaxies to the intergalactic space.

However, it is difficult to measure metal abundance in the cluster outskirts because the ICM density is low and hence 
the X-ray emissions are faint. 
Recently, \citet{Urba17} measured a uniform metalicity ($Z \sim 0.3$ solar) around a half of virial radius across a sample 
of ten nearby massive clusters. 
On the other hand, there have been only a few limited examples of abundance measurements at a virial 
radius for objects with very favored conditions.
\citet{Fuji08} reported $Z \sim 0.2$ solar with a classical abundance table of \citet{Ande89} in the connecting region between A399 and A401,
where the ICM is compressed and hence becomes luminous because of the interaction of the two clusters \citep{Fuji96,Sake04}. 
Their results correspond to $Z \sim 0.3$ solar with a more recent version of an abundance table such as \citet{Lodd03} and \citet{Aspl09},
which is recently confirmed by \citet{Akam17}.
In the Perseus \citep{Wern13} and Virgo cluster \citep{Simi15}, which are the most apparently luminous in X-ray clusters 
and the most nearby cluster, respectively, results equivalent to \citet{Fuji08} on the metal abundance near the virial radius were obtained.
In all cases, relatively high abundance results are reported, which favors the galactic wind scenario.
However, it is clear that we need more samples to investigate whether this trend is universal for galaxy clusters or not.

In the galaxy cluster pair of Abell 3391 and 3395, 
their virial radii overlap with each other, which suggests that
both clusters are in the early phase of a merger and start interacting with each other.
Diffuse X-ray emissions are detected in the region between a A3391 and 3395 with ROSAT and ASCA \citep{Titt01}. 
In addition, Sunyaev-Zel'dovich (SZ) effect \citep{Suny72} signals are also detected in this region with 
Planck \citep{Plan13}.
Actually, this pair is the only example with significant SZ signals from the connecting region except the A399 and 401 pair.
These facts indicate that this system is in a situation quite similar to the A399 and 401 pair and hence appropriate 
for metal abundance measurements near the virial radii.
Though the ICM in the central part of Abell 3395 is investigated in detail \citep{Mark98,Lakh11}, the temperature and abundance 
distributions in the outskirts are not well studied so far.

In this paper, we present Suzaku X-ray observations of the galaxy cluster Abell 3395 including the filament region
connecting to the Abell 3391. The X-ray Imaging Spectrometer (XIS) of Suzaku is suitable for observing 
low surface brightness diffuse sources such as cluster outskirts owing to its low and stable background \citep{Koya07,Mits07}.
The rest of this paper is organized as follows. 
In section 2 we describe the observation and data reduction. In section 3 we present analysis results. 
In section 4 we discuss the results and their implications. In section 5 we summarize the results. 
Canonical cosmological parameters of $H_0 = 70$ Mpc$^{-1}$ km s$^{-1}$ , $\Omega_0 = 0.27$, and $\Lambda_0 = 0.73$ are used 
in this paper. At the mean redshift of Abell 3391 and 3395 ($z=0.05285$), 1 arcmin corresponds to 64.0 kpc.
The solar abundances are normalized to \citet{Aspl09}.
Unless otherwise stated, all uncertainties are given at the 90\% confidence level. 

\section{Observations and Data Reduction}\label{s:obs}

\begin{figure}
  \begin{center}
    \includegraphics[width=8.5cm]{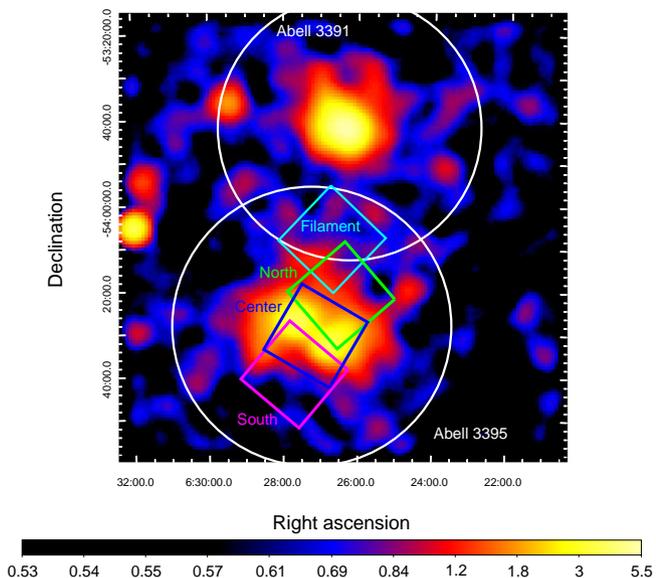} 
  \end{center}
  \caption{ROSAT all sky survey image around Abell 3391 (Northern object) and 3395 (Southern object). 
           The observed XIS fields are displayed with  
           rectangles, and virial radius of each cluster is shown with a white circle.
           Clearly, their virial radii are overlapped with each other at the Filament field.}
           \label{fig:rosat_img}
\end{figure}

We observed a field between Abell 3391 and 3395 with Suzaku on 2014 May 14-15, which we refer as the ``Filament''.
In addition, we used the data of three fields around Abell 3395 in the Suzaku archive, which we refer as the `` North'',
``Center'', and ``South'' from the north to the south. 
Suzaku archive data of the Abell 3404 outskirts field, which is offsetted by 2.5 degree
from the Filament field, are used to estimate the X-ray background (XRB) components. We refer it as ``Background''.
All the observations were performed at XIS nominal pointing. The XIS was operated in the normal full-frame clocking mode. 
The edit mode of the data format was $3 \times 3$ and $5 \times 5$, and combined data of both modes were utilized. 
The spaced-row charge injection was adopted for XIS. 
The calibration data files 20141001 were adopted.
A summary of the observations is given in table \ref{tab:obs}. Figure \ref{fig:rosat_img} shows ROSAT all sky survey image around
Abell 3391 and 3395, where the observed XIS fields are displayed with rectangles. Virial radius of each cluster is also 
shown with a white circle, which is estimated from the mean temperature with an empirical relation \citep{Henry2009},
\begin{equation}
      r_{200}=2.77\pm 0.02 h_{70}^{-1} {\rm Mpc}\frac{\left({kT}/{10 \ {\rm keV}}\right)^{1/2}}{E(z)},
      \label{eq:virial}
\end{equation}
where $h_{70}$ is the Hubble constant normalized by the value $H_0=70$ Mpc$^{-1}$ km s$^{-1}$ and 
$E(z)=[\Omega_{0}(1+z)^3+1-\Omega_{0}]^{1/2}$. $r_{200}$ represents the radius within which the mean density becomes 200 times of
the critical density of the universe. 
Adopting the mean temperature $<kT>=5.39$ and 5.10 keV for Abell 3391 and 3395 \citep{Vikh09}, we obtain 
$r_{200} = 1.98$ and $1.93 h_{70}^{-1}$ Mpc, respectively. Clearly, their virial radii are overlapped with each other at the Filament field.

\begin{table*}
  \caption{Observational log of A3391 and A3395 Field}\label{tab:obs}
  \begin{center}
    \begin{tabular}{cccc} \hline
      Name (Obs.ID)                 & (RA, Dec)        & Observation Date & Exposure (ksec)\footnotemark[$*$] \\ \hline
      Filament (809033010) & (\timeform{96D.691}, \timeform{-54D.123}) & 2014/5/14-15     & 71.8 \\
         North (807031010) & (\timeform{96D.614}, \timeform{-54D.349}) & 2013/2/6-7     & 33.5 \\
        Center (803020010) & (\timeform{96D.784}, \timeform{-54D.507}) & 2008/12/15-17     & 81.2 \\
         South (807032010) & (\timeform{96D.927}, \timeform{-54D.657}) & 2013/2/16-17     & 39.3 \\
    Background (804089010) & (\timeform{101D.409}, \timeform{-54D.203}) & 2009/5/20-23    & 87.8 \\ \hline 
   \multicolumn{2}{@{}l@{}}{\hbox to 0pt{\parbox{180mm}{\footnotesize
       \footnotemark[$*$] An effective exposure time after screening as described in the text.
     }\hss}}
    \end{tabular}
  \end{center}
\end{table*}

\begin{figure}
  \begin{center}
    \includegraphics[width=8.5cm]{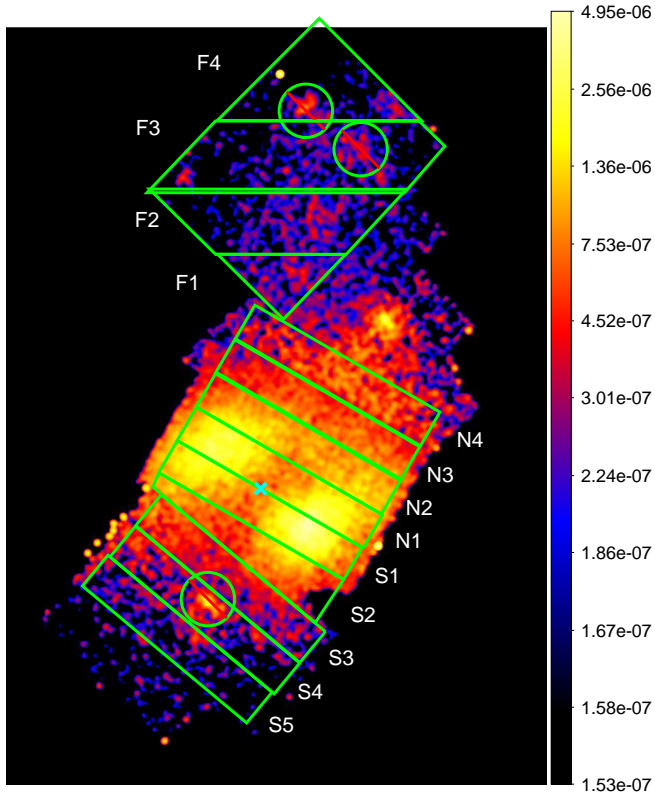} 
  \end{center}
  \caption{An XIS3 image of Abell 3395 and filament region in the 0.5-8.0 keV band, which was corrected for exposure 
           after subtracting NXB, and smoothed with Gaussian kernel with $\sigma=0.26'$. The region
           utilized to investigate temperature and abundance profiles in section \ref{s:an_re} are displayed 
           as green boxes. The region names are also shown in white. Green circles are excluded regions of a point source.}
           \label{fig:suzaku_img}
\end{figure}

The XIS data were processed through standard criteria as follows. 
Events with a GRADE of 0, 2, 3, 4, 6 and STATUS with 0:524287 were 
extracted. Data obtained at the South Atlantic Anomaly (SAA), within 436 s after the passage of SAA, 
and at low elevation angles from an Earth rim of $< 5^{\circ}$ and a Sun-lit Earth rim of $< 20^{\circ}$ were excluded. 
In addition to the above-mentioned criteria, we performed event screening with cut-off rigidity greater than 8 GV.
For the data of Filament, North, and South, we applied additional processing for XIS1 to reduce the non X-ray background (NXB) level, 
which increased after changing the amount of charge injection, following descriptions in the XIS analysis topics 
(see \verb|http://www.astro.isas.jaxa.jp/suzaku/analysis/xis/|  \verb|xis1_ci_6_nxb/|).
NXB spectra and images of XIS were generated using the ftool ``xisnxbgen'' \citep{Tawa08}. 
For all XIS CCD chips, fields corresponding to the damaged part of XIS0 CCD chip because of a micrometeorite accident
were not utilized in the following analysis. 
Figure \ref{fig:suzaku_img} represents an 0.5-8.0 keV XIS3 image, which was corrected for exposure  
after subtracting NXB, and smoothed it by a Gaussian kernel with $\sigma=0.26'$.
The regions utilized to investigate surface brightness, temperature, and abundance profiles in section \ref{s:an_re} are 
displayed as green boxes. The region names are also shown in white. Green circles are excluded regions of a point source.

\section{Analysis and Results}\label{s:an_re}
To analyze the XIS data, we generated redistribution matrix files (RMFs) with ftool ``xisrmfgen'' 
and ancillary response files (ARFs), which describe the response of the X-ray telescope aboard Suzaku
and amount of the XIS optical blocking filters contamination, with ``xissimarfgen'' \citep{Ishi07}.
To generate an ARF, we utilized uniform emission from a circular region with 20' radius as an input image.
The energy range near the Si edge (1.7-1.8 keV) was ignored to avoid uncalibrated structures.

\subsection{The X-ray Background}
It is necessary to estimate an appropriate XRB model from astrophysical origin in addition to the NXB.
Because all of the FOVs of North, Center, South and Filament field are expected to be filled with 
the ICM emission, we used the Background field data, which were originally taken for Abell 3404 outskirts
investigation. Because the center of Abell 3404 is near the east end of the FOV, we used only a western part 
outside the virial radius of the cluster to determine the XRB model. We consider the cosmic X-ray background (CXB),
the Milky Way halo's hot gas (MWH) and the local hot bubble (LHB) as the XRB component. Then, the spectrum was fitted
with the following model:
\begin{eqnarray}
  {\rm apec}_{\rm LHB} + {\rm wabs} \times ( {\rm apec}_{\rm MWH} + {\rm powerlaw}_{\rm CXB} ),
\end{eqnarray}
where ${\rm apec}_{\rm LHB}$, ${\rm apec}_{\rm MWH}$, and ${\rm powerlaw}_{\rm CXB}$ represent the LHB, MWH, and CXB, respectively.
The temperatures of the LHB and MWH were fixed to be 0.08 and 0.3 keV, respectively. The metal abundance and redshift of both
LHB and MWH were also fixed to be solar and zero, respectively. The powerlaw index of the CXB was fixed to be 1.4 \citep{Kush02}.
We assumed $N_{\rm H} = 6.35 \times 10^{20}$cm$^{-2}$ for the Galactic absorption \citep{Dick90}.
The energy band of 0.7-6.0 keV was used for the spectral fitting of the background data.

\begin{figure}
  \begin{center}
    \includegraphics[width=6cm, angle=270]{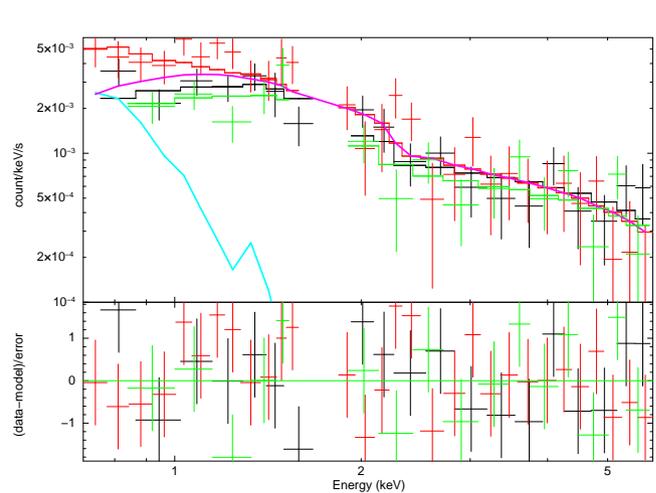} 
  \end{center}
  \caption{The XIS spectra of the background field fitted with the XRB model described in the text,
          where the black, red, and green crosses show the spectra of XIS0, XIS1, and XIS3, respectively.
          MWH and CXB components of the best fit model are also plotted as light-blue and magenta histograms, 
          respectively. The LHB component is not seen because its contribution is too weak and limited 
          in the very low energy range.}
          \label{fig:bgdspec}
\end{figure}

Figure \ref{fig:bgdspec} shows the background field spectra fitted with the above-mentioned model, 
where the black, red, and green crosses show the spectra of XIS0, XIS1, and XIS3, respectively.
MWH and CXB components of the best fit model are also plotted as light-blue and magenta histograms, respectively.
The LHB component is not seen because its contribution is too weak and limited in the very low energy range.
The results of the fit are summarized in table \ref{tab:bgd}. In general the background spectra are well fitted
with the model used here.

\begin{table*}
  \caption{Best-fit parameters for the XIS spectra of the background field}\label{tab:bgd}
  \begin{center}
    \begin{tabular}{ccc} \hline 
      Model Component   & Parameter                          & Value
     \\ \hline
      LHB               & $kT$\footnotemark[$*$]             & 0.08 (fixed) \\
                        & $N$\footnotemark[$\dagger$]        & $1.63^{+1.78 \times 10^4}_{-0.74} \times 10^{-5}$ \\  \hline
      MWH               & $kT$\footnotemark[$*$]             & 0.3  (fixed) \\
                        & $N$\footnotemark[$\dagger$]        & $5.57^{+2.38}_{-3.70} \times 10^{-4}$  \\ \hline
      CXB               & $\Gamma$\footnotemark[$\ddagger$]  & 1.4 (fixed) \\
                        & $N$\footnotemark[$\S$]             & $8.79^{+0.72}_{-0.72} \times 10^{-4}$  \\ \hline
                        & $\chi^2/d.o.f$                     & 53.08/62 \\ \hline
    \\
   \multicolumn{2}{@{}l@{}}{\hbox to 0pt{\parbox{180mm}{\footnotesize
       \footnotemark[$*$] Temperature of the each component in keV.
       \par\noindent
       \footnotemark[$\dagger$] Normalization in the ${\rm apec}$ code for each component scaled with a factor $1/400 \pi$. \\
                     $N=(1/400 \pi) \int n_{\rm e} n_{\rm H} dV / [ 4 \pi (1+z)^2 D_{\rm A}^2 ] \times 10^{-14}$ cm$^{-5}$ arcmin$^{-2}$,\\
                                where $D_{\rm A}$ is the angular diameter distance to the source.
       \par\noindent
       \footnotemark[$\ddagger$] Photon index of the power-law component.
       \par\noindent
       \footnotemark[$\S$] Normalization in the power-law component in photons keV$^{-1}$ cm$^{-2}$ s$^{-1}$ at 1 keV
     }\hss}}
    \end{tabular}
  \end{center}
\end{table*}

\subsection{X-ray Surface Brightness Profile}
In order to confirm the detection of the ICM emission in the outskirts of Abell 3395, we made a surface brightness profile
from 0.5-2.0 keV XIS3 mosaic image and compared it with those of XRB and NXB in a way similar to \citet{Kawa10}. 
First, we made a raw (including XRB and NXB) 0.5-2.0 keV surface brightness profile
from the southern outskirt of Abell 3395 to the north filament with regions shown in figure \ref{fig:suzaku_img}. We also created
a corresponding 0.5-2.0 keV NXB profile from the NXB mosaic image. Adopting the XRB model (LHB+MWH+CXB) described in the previous subsection, 
we simulated an XRB mosaic image through the ftool ``xissim'' with 100 ks exposures. Finally, we obtained a background-subtracted profile, 
which should be a contribution from the ICM emission. For the region N3, the data of the North and Center fields were analyzed independently.
For the regions S2 and S3, the data were also analyzed independently for the Center and South fields.

\begin{figure}
  \begin{center}
    \includegraphics[width=6.5cm, angle=270]{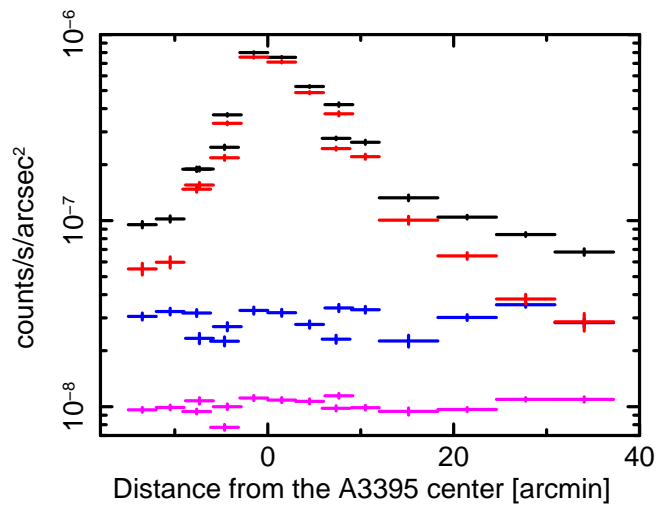} 
  \end{center}
  \caption{
           The surface brightness profiles from the southern outskirt of A3395 to the north filament. The raw, XRB, and NXB profiles are shown in 
           black, blue, and magenta crosses, respectively. The background-subtracted profile is also shown in red crosses. 
           The horizontal axis shows the angular distance from the A3395 center, where the north and south directions are positive and negative, 
           respectively.
           }
          \label{fig:srbp}
\end{figure}

Figure \ref{fig:srbp} shows the raw, XRB, and NXB profiles with black, blue, 
and magenta crosses, respectively. 
The background-subtracted profile is also shown in red crosses. The horizontal axis shows the angular distance from 
the A3395 center, where the north and south directions are positive and negative, respectively. Note that these profiles are not corrected
for the vignetting effect. Thus, apparently inconsistent results are seen for the data from the same sky region but different observed fields.
However, this does not matter because our main purpose is comparing the ICM signals to the XRB and NXB.
It is clear that even in the faintest region the ICM emission is comparable to the XRB and significantly more than the NXB.

\subsection{Temperature Measurements}
The temperature structure along the large scale structure filament including A3391and 3395 is investigated. The regions for this analysis 
is displayed as green boxes in figure \ref{fig:suzaku_img}. We fit the spectrum of each region by the following model:
\begin{eqnarray}
     {\rm const} \times \{ {\rm apec}_{\rm LHB} + {\rm wabs} \times ( {\rm apec}_{\rm MWH} + {\rm powerlaw}_{\rm CXB} \nonumber \\
               + {\rm apec}_{\rm ICM} ) \},
\end{eqnarray}
where, ${\rm apec}_{\rm ICM}$ represents the emission from the ICM. We fixed all the parameters of the XRB components to the values
determined from the background field analysis. 
We assumed that $N_{\rm H}=6.13 \times 10^{20}$ cm$^{-2}$ for the Galactic absorption \citep{Dick90}. 
We also assumed that $z=0.05285$ for the ICM in regions of the Filament field (region F1-F4), 
which is a mean value of those of Abell 3391 and 3395.
On the other hand, $z=0.0515$, which is a mean of those of double peaks of Abell 3395, was assumed for the ICM in the other regions.
In the fitting of regions F2, F3, F4, S4, and S5, abundance of the ICM component was fixed to 0.3 solar, following the former results
about abundance measurements in the cluster outskirts \citep{Fuji08, Wern13, Simi15}, while it was allowed 
to vary freely for the other regions.
A parameter ${\rm const}$ is introduced to take into account of possible calibration uncertainties among different XIS sensors. It was fixed
to be unity for XIS1 and allowed to vary freely for XIS0 and XIS3. The other parameters were common among all XIS sensors.
The energy band of 0.7-7.0 keV was used for the spectral fitting to measure the temperature of the ICM.
Systematic errors of CXB were estimated in the same way as in \citet{Itah15}. We adopted an upper cutoff flux of 
$1 \times 10^{-13}$ erg cm$^{-2}$ s$^{-1}$ considering the spectral analysis results of the visually inspected point sources in the filament
field (see Appendix). We also estimated NXB systematic errors from the reported reproducibility (4.9\% at 90 \% confidence level) 
by \citet{Tawa08}. For the region N3, spectra extracted from the North and Center field data were analyzed separately.
For the regions S2 and S3, spectral analyses were also performed separately for the data of the Center and South fields.

\begin{figure}
  \begin{center}
    \includegraphics[width=8.5cm]{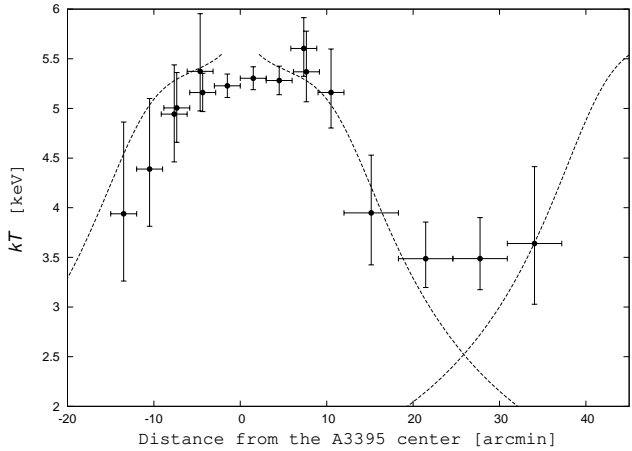} 
  \end{center}
  \caption{The temperature profile of Abell 3395 from the south outskirts to the north filament toward A3391.
           The horizontal axis shows the angular distance from the A3395 center, where the north and south directions are 
           positive and negative, respectively. Universal temperature profiles proposed by \citet{Okab14} are also plotted 
           for Abell 3391 and 3395 by dashed lines.}
          \label{fig:temp_dist}
\end{figure}

The results of the fitting for each region are listed in table \ref{tab:temp_fit}.
Figure \ref{fig:temp_dist} shows the temperature profile of Abell 3395 from the south outskirts to the north filament toward A3391.
The horizontal axis shows the angular distance from the A3395 center, where the north and south directions are positive and negative,
respectively. Universal temperature profiles proposed by \citet{Okab14} are also plotted for Abell 3391 and 3395 by dashed lines, which will
be discussed in subsection \ref{ss:temp_dist} in more detail.

\begin{table*}[htbp]
  \caption{Fitting results for temperature measurements}\label{tab:temp_fit}
  \begin{center}
    \begin{tabular}{cccc}
      \hline \hline
      Region Number & $kT$ (keV)\footnotemark[$*$] &  $N$ ($\times$ $10^{-2}$)\footnotemark[$* \dagger$] & $\chi^2/d.o.f$ \\ \hline
            F1      & $3.95^{+0.58+0.46+0.11 }_{-0.52-0.61-0.12 }$  &  $1.52^{+0.17+0.15+0.00 }_{-0.16-0.15-0.00 }$  &   88.11/90 \\
            F2      & $3.49^{+0.37+0.45+0.15 }_{-0.29-0.41-0.14 }$  &  $0.90^{+0.05+0.09+0.01 }_{-0.05-0.09-0.00 }$  &   266.09/240 \\
            F3      & $3.49^{+0.41+0.46+0.19 }_{-0.31-0.42-0.16 }$  &  $0.80^{+0.04+0.09+0.01 }_{-0.04-0.09-0.01 }$  &   233.94/237 \\
            F4      & $3.64^{+0.77+0.60+0.23 }_{-0.61-0.67-0.27 }$  &  $0.86^{+0.08+0.13+0.01 }_{-0.08-0.13-0.01 }$  &   107.42/82  \\
            N1      & $5.30^{+0.12+0.04+0.02 }_{-0.11-0.04-0.02 }$  &  $11.80^{+0.17+0.12+0.01 }_{-0.17-0.12-0.01 }$ &  1280.74/1184 \\
            N2      & $5.28^{+0.14+0.06+0.03 }_{-0.14-0.06-0.03 }$  &  $8.91^{+0.16+0.12+0.01 }_{-0.16-0.12-0.01 }$  &   876.71/868 \\
            N3ce\footnotemark[$\ddagger$]    & $5.60^{+0.31+0.12+0.07 }_{-0.28-0.12-0.08 }$  &  $5.40^{+0.15+0.12+0.01 }_{-0.15-0.12-0.02 }$  &   463.20/480 \\
            N3no\footnotemark[$\ddagger$]    & $5.37^{+0.41+0.11+0.04 }_{-0.30-0.09-0.04 }$  &  $5.64^{+0.22+0.11+0.01 }_{-0.21-0.11-0.01 }$  &   289.04/245 \\
            N4      & $5.16^{+0.44+0.14+0.05 }_{-0.36-0.15-0.06 }$  &  $3.56^{+0.16+0.12+0.01 }_{-0.16-0.12-0.01 }$  &   199.35/201 \\
            S1      & $5.23^{+0.12+0.04+0.02 }_{-0.12-0.05-0.02 }$  &  $11.32^{+0.18+0.11+0.01 }_{-0.18-0.12-0.01 }$ &  1161.15/1090 \\
            S2ce\footnotemark[$\ddagger$]    & $5.16^{+0.19+0.08+0.04 }_{-0.19-0.09-0.04 }$  &  $6.46^{+0.16+0.13+0.01 }_{-0.16-0.13-0.01 }$  &   545.18/535  \\
            S2so\footnotemark[$\ddagger$]    & $5.37^{+0.58+0.18+0.10 }_{-0.40-0.12-0.06 }$  &  $4.91^{+0.23+0.13+0.00 }_{-0.22-0.13-0.00 }$  &   167.30/171 \\
            S3ce\footnotemark[$\ddagger$]    & $5.00^{+0.36+0.16+0.10 }_{-0.35-0.17-0.11 }$  &  $3.49^{+0.14+0.12+0.01 }_{-0.14-0.12-0.01 }$  &   202.70/256 \\
            S3so\footnotemark[$\ddagger$]    & $4.94^{+0.50+0.23+0.09 }_{-0.48-0.25-0.10 }$  &  $2.34^{+0.14+0.12+0.00 }_{-0.14-0.12-0.01 }$  &   165.27/141 \\
            S4      & $4.39^{+0.71+0.42+0.14 }_{-0.58-0.45-0.19 }$  &  $1.46^{+0.10+0.13+0.01 }_{-0.10-0.13-0.01 }$  &    106.56/90 \\
            S5      & $3.94^{+0.92+0.70+0.20 }_{-0.68-0.70-0.23 }$  &  $0.71^{+0.07+0.11+0.01 }_{-0.07-0.11-0.01 }$  &     61.41/77 \\
      \hline \\
      \multicolumn{2}{@{}l@{}}{\hbox to 0pt{\parbox{180mm}{\footnotesize
      \footnotemark[$*$] The first, second, and third errors are statistical, CXB systematic, and NXB systematic, respectively.
      \par\noindent
      \footnotemark[$\dagger$] Normalization in the ${\rm apec}$ code scaled in the same way as in table \ref{tab:bgd}. 
      \par\noindent
      \footnotemark[$\ddagger$] Spectra extracted from the North, Center, and South are 
                                denoted as no, ce, and so, respectively 
     }\hss}}
    \end{tabular}
  \end{center}
\end{table*}

\subsection{Abundance Measurements}
The abundance profile is also investigated along the filament structure. In order to determine the abundance, 
we fit the spectrum of each region in a way similar to the previous subsection, but using the energy range of
2.0-7.0 keV.
Metal abundance is determined through the iron L ($\sim$ 1 keV) and K ($\sim$ 6.7 keV) lines in the spectral fitting of hot thin thermal 
plasma such as ICM. However, we should take special care of abundance determination through the iron L lines \citep{Sasa15,Simi15}. 
For example, disagreement
among plasma emission models is relatively prominent in the energy range near iron L lines, and uncertainties of the Galactic 
background components (LHB and MWH) could seriously affect the fitting results. In addition, multi-temperature
structures could cause biases in measuring the metalicities. These are especially true when
the emissions from the concerned objects are relatively faint in the energy near the iron K lines.
Therefore, we ignored lower energy band around the L lines
and measured the metal abundance using only the K lines. In the abundance measurements, the regions F1, F2, F3, and F4 were merged 
to improve the statistics. The region S4 and S5 were also treated in the same way. Systematic errors of CXB and NXB were estimated in 
the same way as in the previous subsection.

\begin{figure}
  \begin{center}
    \includegraphics[width=6cm, angle=270]{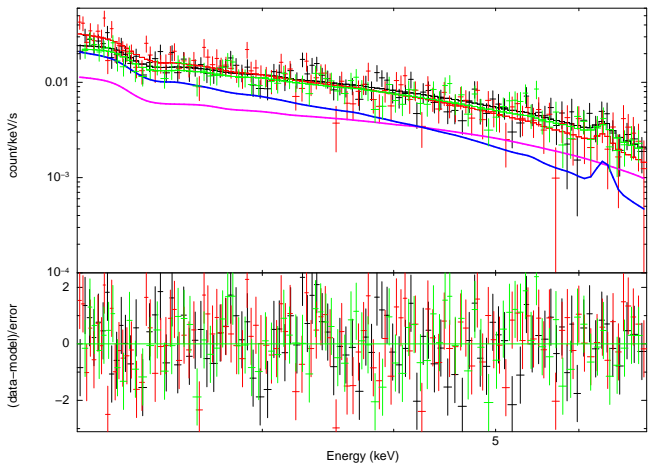} 
  \end{center}
  \caption{The XIS spectra of the Filament field (F1-4) fitted with the model described in the text 
           using the energy range of 2.0-7.0 keV,
          where the black, red, and green crosses show the spectra of XIS0, XIS1, and XIS3, respectively.
          ICM and CXB components of the model are also plotted as blue and magenta histograms, 
          respectively.}
          \label{fig:f1-4_spec}
\end{figure}

\begin{figure}
  \begin{center}
    \includegraphics[width=8.5cm]{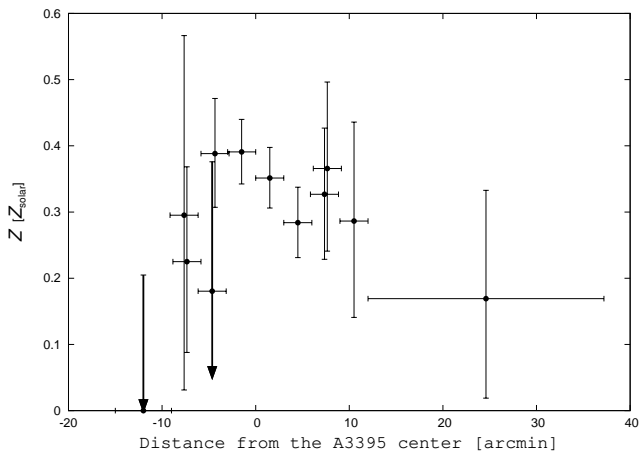} 
  \end{center}
  \caption{The abundance profile of Abell 3395 from the south outskirts to the north filament toward toward A3391.
           The horizontal axis shows the angular distance from the A3395 center, 
           where the north and south directions are positive and negative, respectively.}
          \label{fig:abd_dist}
\end{figure}

The obtained results are listed in table \ref{tab:abun_fit}. The XIS spectra of the Filament field (F1-4) fitted 
through the above-mentioned procedure are shown in figure \ref{fig:f1-4_spec}, where the black, red, and green crosses 
show the spectra of XIS0, XIS1, and XIS3, respectively. ICM and CXB components of the model are also plotted 
as blue and magenta histograms, respectively. In the energy range near Fe-K lines, ICM component is comparable to the CXB.
Figure \ref{fig:abd_dist} shows the abundance profile of Abell 3395 from the south outskirts to the north filament toward toward A3391.
Again, the horizontal axis shows the angular distance from the A3395 center, where the north and south directions are positive and negative,
respectively. In the filament region, we get $Z \sim 0.17$ solar though the errors are large. In the southern outskirt, on the other hand,
only an upper limit ($Z<0.205$ solar) is obtained because of the poor statistics.

To check how including Fe-L lines in the fit affects the metalicity measurements, we fit the spectrum of the 
filament region data (F1-4) in the same way, but utilizing the energy range of 0.7-7.0 keV. The result becomes
$Z=0.032^{+0.079+0.027+0.028}_{-0.032-0.018-0.016}$ solar, which means that we get only an upper limit $(Z<0.120)$.
Therefore, including Fe-L lines results in serious underestimation of the metal abundance.

\begin{table*}[htbp]
  \caption{Abundance determined through Fe K lines}\label{tab:abun_fit}
  \begin{center}
    \begin{tabular}{ccc}
      \hline \hline
      Region Number &  $Z$ ($Z_{\odot}$)\footnotemark[$*$]  & $\chi^2/d.o.f$  \\ 
      \hline 
       F1-4         & $0.169^{+0.164+0.009+0.018 }_{-0.150-0.004-0.015 }$    &    300.19/295  \\
       N1           & $0.351^{+0.046+0.001+0.002 }_{-0.045-0.001-0.003 }$    &    603.22/584  \\
       N2           & $0.284^{+0.054+0.003+0.002 }_{-0.053-0.003-0.002 }$    &    397.19/421  \\
       N3ce\footnotemark[$\dagger$]         & $0.327^{+0.100+0.003+0.006 }_{-0.098-0.003-0.006 }$    &    209.66/221  \\
       N3no\footnotemark[$\dagger$]         & $0.366^{+0.131+0.007+0.002 }_{-0.125-0.006-0.003 }$    &    139.70/109  \\
       N4           & $0.286^{+0.149+0.007+0.004 }_{-0.145-0.012-0.004 }$    &     92.30/90   \\
       S1           & $0.391^{+0.049+0.004+0.001 }_{-0.048-0.004-0.001 }$    &    563.34/546  \\
       S2ce\footnotemark[$\dagger$]         & $0.388^{+0.083+0.007+0.002 }_{-0.081-0.007-0.003 }$    &    271.01/247  \\
       S2so\footnotemark[$\dagger$]         & $<0.376$    &      71.28/73  \\
       S3ce\footnotemark[$\dagger$]         & $0.225^{+0.143+0.004+0.008 }_{-0.137-0.004-0.009 }$    &      87.76/116 \\
       S3so\footnotemark[$\dagger$]         & $0.295^{+0.271+0.003+0.015 }_{-0.260-0.003-0.016 }$    &      57.49/62  \\
       S4-5         & $<0.205$  &  95.69/82 \\
  \hline \\
      \multicolumn{2}{@{}l@{}}{\hbox to 0pt{\parbox{180mm}{\footnotesize
      \footnotemark[$*$] The errors are represented in the same way as in table \ref{tab:temp_fit}.
      \par\noindent
      \footnotemark[$\dagger$] Spectra extracted from the North, Center, and South are 
                                denoted as no, ce, and so, respectively 
     }\hss}}
  \end{tabular}
  \end{center}
\end{table*}

\section{Discussion}

\subsection{Temperature Distribution}\label{ss:temp_dist}
We compare the obtained temperature results with a universal temperature profile proposed by \citet{Okab14}
for relaxed galaxy clusters with lensing and X-ray data. The profile is 
\begin{equation}
        \frac{kT}{kT_{*}}=\biggl( \tilde{r}_0^{-1} \frac{r}{r_{200}} \biggr)^{\frac{3}{5}\alpha - \frac{2}{5}\gamma} 
                  \biggr[ 1 + ( \tilde{r}_0^{-1} \frac{r}{r_{200}} )^{\beta}  \biggl]^{-(\frac{2}{5}\delta - \frac{2}{5}\gamma + \frac{3}{5} \alpha)/\beta},
\end{equation}
where $\alpha=1.16^{+0.17}_{-0.12}$, $\beta=5.52^{+2.87}_{-2.64}$, $\gamma=1.82^{+0.28}_{-0.30}$, $\delta=2.72^{+0.34}_{-0.35}$, and 
$\tilde{r}_0 = 0.45^{+0.08}_{-0.07}$.  The normalization factor of the temperature is
\begin{equation}
      kT_{*} = 1.27^{+0.24}_{-0.19} {\rm keV} \biggr[ \frac{M_{200} E(z)}{10^{14} h^{-1}_{70} M_{\odot}} \biggl]^{2/3},
\end{equation}
where $M_{200}$ is the total mass inside $r_{200}$. Though it is ideal that both $M_{200}$ and $r_{200}$ should be determined through 
direct mass profile measurements such as weak gravitational lensing technique, we do not have available data for these clusters. 
Instead, we utilize $r_{200}$ derived from the empirical relation of equation (\ref{eq:virial}) as described in section \ref{s:obs}.
We calculate $M_{200}$ simply assuming that the mean density inside $r_{200}$ is 200 times of the critical density 
of the universe. 
As a result, $M_{200}= 8.84 \times 10^{14} h_{70}^{-1} M_{\odot}$ and $8.19 \times 10^{14} h_{70}^{-1} M_{\odot}$ for A3391 and A3395, respectively.
A comparison of the measured temperature (crosses) with the profile of \citet{Okab14} (dashed lines) is shown in figure \ref{fig:temp_dist}.
Basically, the obtained temperature results agree well with the universal profile except for the filament regions.
While southern part is well matched, marginally higher temperatures are seen in the northern 
interconnecting region. This might be a hint of the interaction of Abell 3395 and 3391.

\subsection{Abundance Distribution}
As mentioned in the introduction, heavy-element abundance in the cluster outskirts contains crucial information
on transport processes of the heavy elements from galaxies to the intergalactic space. In the previous studies about this issue
\citep{Fuji08, Wern13, Simi15}, relatively high abundance ($Z \sim 0.3$ solar) are reported at the virial radius, which
suggests early enrichment of heavy elements owing to the galactic wind scenario. 
We detected slightly lower abundance in the filament region (see figure \ref{fig:abd_dist} and table \ref{tab:abun_fit}), 
but the value is consistent with the previous results considering the errors. Therefore, we can say our results are consistent
with the early enrichment scenario. However, we cannot rule out the significantly lower abundance in the filament region.
if this is true, it could indicate diversity of metal enrichment history in clusters. 
Future deeper observations are desired to measure abundance more accurately and clarify this possibility. 
As for the southern outskirt, we did not detect significant
metal abundance, which is not so surprising considering the poor statistics, and
obtained only an upper limit of $Z < 0.205$. This also suggests slightly lower abundance than the previous results.
Again, future deeper observations are necessary to detect and determine the abundance and compare the results with the northern filament region.

\subsection{Comparison with Planck Results}
The SZ effect intensity is characterized by the Compton $y$ parameter,
\begin{eqnarray}
  y = \int \biggl( \frac{k T_{\rm e}}{m_{\rm e} c^2} \biggr) n_{\rm e} \sigma_{\rm T}  dl,
\end{eqnarray}
where $\sigma_{\rm T}$ is the Thomson cross section and the integration is performed along the line-of-sight.
SZ effect signals are detected in the interconnecting region between A3391 and A3395 with Planck satellite \citep{Plan13} where 
the minimum of Compton $y$ parameters is $\sim 7 \times 10^{-6}$
 with typical errors of $\sim 1 \times 10^{-6}$ as seen in figure 2 of \citet{Plan13}. 
We estimate the $y$ parameters independently for the connecting region through
X-ray results alone to compare the Planck results.  We estimated electron number densities from the normalization
of the apec code shown in table \ref{tab:temp_fit}. 
Considering that a diagonal line of the Filament field is nearly parallel to the interconnecting filament and that the XIS field of view is
a $17.8' \times 17.8'$ square,
we assume that the emitting volume is a cylinder whose radius and height are half of and equal to $\sqrt{2} \times 17.8'$ on the sky plain, 
respectively. In other words, the connecting region is assumed to be a cylinder whose radius and height are 806.5 kpc and 1613 kpc, 
respectively, and its bottom surfaces are assume to be perpendicular to the sky plane.
The obtained electron number densities are listed in table \ref{tab:density}. 
Using these results and temperatures in table \ref{tab:temp_fit}, the Compton
$y$ parameters are calculated and listed in table \ref{tab:ypara}, which are lower than the Planck results.  

One main uncertainty is the assumed geometry of the emitting region to derive electron densities. 
The normalization of the apec code $N$ is proportional to $n_{\rm e}^2 L$, where $L$ is the length of the line-of-sight direction. 
On the other hand, the Compton $y$ parameter is proportional to $n_{\rm e} L$.
Thus, $y \sim L^{1/2}$ given $N$. This may indicate that the line-of-sight length of the filament is larger than what we assumed. 
Alternatively, if the filament is inclined to the line-of-sight direction, effective length of the line-of-sight direction becomes longer
and a similar situation is realized. 
If the discrepancy between the SZ results and ours is fully attributed to this geometrical effect, the actual line-of-sight length should
be $\sim 5$ times larger than the assumed one. If we explain this factor with an inclined filament, the filament should be inclined
to the line-of-sight direction by $\sim 10$ degree. 
Therefore, though a part of the discrepancy would be attributed to the assumed geometry, 
the difference seems too large to be explained by the geometrical factor alone. 

Another possible explanation of the discrepancy relies on temperature measurements. The mean X-ray temperature could be underestimated
due to unresolved multi-temperature structures. Or, the spectral fits could miss some hot X-ray emissions of the shock heated gas between
the merging clusters, due to the background uncertainties in the hard band. Such underestimation of the temperature
would also result in a lower Compton $y$ parameter derived from the X-ray data.

\begin{table}[htbp]
  \caption{Electron number density in the filament}\label{tab:density}
  \begin{center}
    \begin{tabular}{cc}
      \hline \hline
      Region Number & $n_{\rm e}$ ($10^{-4}$ ${\rm cm^{-3}}$)\footnotemark[$*$] \\ \hline
           F1       & $1.67^{+0.09+0.08+0.00}_{-0.08-0.07-0.00}$ \\
           F2       & $1.28^{+0.03+0.07+0.01}_{-0.03-0.06-0.01}$ \\
           F3       & $1.21^{+0.03+0.07+0.01}_{-0.03-0.06-0.01}$ \\
           F4       & $1.25^{+0.06+0.10+0.01}_{-0.06-0.09-0.01}$ \\
      \hline \\
      \multicolumn{2}{@{}l@{}}{\hbox to 0pt{\parbox{180mm}{\footnotesize
      \footnotemark[$*$] The errors are represented in the same way as in table \ref{tab:temp_fit}.
     }\hss}}
    \end{tabular}
  \end{center}
\end{table}

\begin{table}[htbp]
  \caption{Compton $y$ parameters in the filament}
  \begin{center}
    \begin{tabular}{cc}
      \hline \hline
      Region Number & $y$ ($10^{-6}$)\footnotemark[$*$] \\ \hline
       F1 & $4.28_{-0.60-0.68-0.13}^{+0.67+0.54+0.12}$  \\
       F2 & $2.90_{-0.25-0.37-0.12}^{+0.31+0.41+0.13}$  \\
       F3 & $2.74_{-0.25-0.36-0.13}^{+0.33+0.39+0.15}$  \\
       F4 & $2.95_{-0.51-0.58-0.22}^{+0.64+0.54+0.19}$  \\
      \hline \\
      \multicolumn{2}{@{}l@{}}{\hbox to 0pt{\parbox{180mm}{\footnotesize
      \footnotemark[$*$] The errors are represented in the same way as in table \ref{tab:temp_fit}.
     }\hss}}
      \label{tab:ypara}
    \end{tabular}
  \end{center}
\end{table}

\section{Conclusions}
We observed the region between the Abell 3395 and 3319 with Suzaku. Using the both obtained and archive data of Abell 3395 field,
we made temperature and abundance profiles of Abell 3395 out to the $r_{200}$ along the large scale structure filament including Abell 3391.
The obtained temperature profile is consistent with the universal temperature profile of \citet{Okab14} except for the filament regions,
where a hint of the temperature increase is seen. This might be due to the interaction of Abell 3391 and 3395 
in the early phase of a merger. 
Our results about metal abundance in the filament region are consistent with the previous results of other clusters 
\citep{Fuji08, Wern13, Simi15} within errors, though the value is slightly lower.
Therefore, our results are consistent with the early enrichment scenario though errors are large. Future deeper observations are 
desired to measure the metal abundance more precisely and clarify the enrichment scenario.

We derived Compton $y$ parameters in the region between A3391 and 3395 only from the X-ray observables assuming a simple geometry. 
The obtained $y$ parameters are smaller than SZ results with Planck \citep{Plan13}. 
This discrepancy would be attributed to a more elaborate geometry such as a filament inclined to the line-of-sight direction,
or underestimation of the X-ray temperature because of the unresolved multi-temperature structures or undetected hot X-ray
emission of the shock heated ICM.

\begin{ack}
The authors would like to thank K. Sato, K. Matsushita, S. Shibata, and H. Ohno for valuable comments. 
We also thank the Suzaku operation team for their support in planning and executing this observation.
This work is supported in part by Japan Society for the Promotion of Science (JSPS) KAKENHI Grant Number 26400218(MT)
and 15K05080 (YF). 
H.A acknowledges the support of NWO via a Vein grant.  SRON is supported financially by NWO, 
the Netherlands Organization for Scientific Research. 
\end{ack}

\appendix 
\section*{Point Sources}
We investigate the X-ray spectra of two point sources in the filament field, which are recognized by visual inspection.
We refer the north-eastern and south-western sources as point sources A and B, respectively. The spectra of these sources
were extracted from the circular region with 2' radius centered on the sources taking into account of the spatial resolution 
of the X-ray telescope of Suzaku. The background spectra were constructed from the data 
in the 9.1' $\times$ 11.7' rectangular region around the source regions, which should contain both NXB and astrophysical origin background
components such as LHB, MWH, CXB, and ICM. We tried to fit the source spectra after subtracting the background
with a model of ${\rm const} \times {\rm wabs} \times {\rm powerlaw}$ or ${\rm const} \times {\rm wabs} \times {\rm apec}$.
As for the Galactic absorption, we adopted $N_{\rm H} = 6.13 \times 10^{20}$ cm$^{-2}$ \citep{Dick90}.

\begin{figure}
  \begin{center}
    \includegraphics[width=6cm, angle=270]{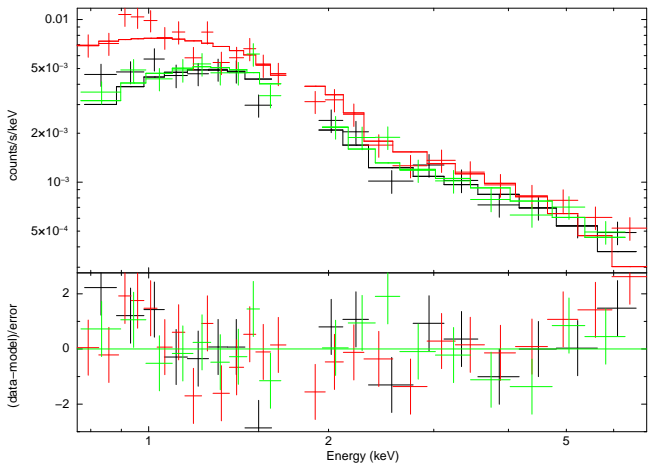} 
  \end{center}
  \caption{The Spectra of the point source A fitted with the power-law model, 
           where the black, red, and green crosses show the spectra of XIS0, XIS1, and XIS3, respectively.
           The spectra of best fit model are also plotted as sold histograms.}
          \label{fig:ps_A_pow}
\end{figure}

The spectra of the source A are better fitted with the powerlaw model than apec. Figure \ref{fig:ps_A_pow}
shows the source A spectra fitted with the power-law model and the best-fit parameters are listed in table \ref{tab:ps_A_pow}.
The photon index obtained from the fitting results is within a range of those of typical AGNs.
The flux of the source in the energy band of 2.0-10.0 keV obtained from each XIS sensor is listed in table \ref{tab:ps_A_flux}, 
which indicates $\sim 3 \times 10^{-13}$  erg s$^{-1}$ cm$^{-2}$.
Using the NASA Extragalactic Database (NED), we confirm that the point source A coincides in position with the galaxy 1WGA J026.8-5402. 
Its X-ray flux in the 0.2-2.0 keV band obtained by ROSAT is $(6.26 \pm 1.80) \times 10^{-14}$ erg s$^{-1}$ cm$^{-2}$, which is converted into 
$(8.06 \pm 2.32) \times 10^{-14}$ erg s$^{-1}$ cm$^{-2}$ in the energy band of 2.0-10.0 keV. Therefore, the flux of our results
is a few times larger than that with ROSAT, which is not so surprising considering that time variability is naturally expected in AGNs.

\begin{table}[htbp]
  \caption{Fitting results of the point source A spectra with the power-law model}
  \begin{center}
    \begin{tabular}{ccc}
      \hline \hline
      model component & parameter & value  \\ \hline
      powerlaw & $\Gamma$\footnotemark[$*$] & $1.67_{-0.06}^{+0.06}$  \\
               & $N$\footnotemark[$\dagger$] & $7.49_{-0.47}^{+0.48}$ $\times$ $10^{-5}$ \\ \hline
               & $\chi^2/d.o.f$ & 70.23/57  \\
      \hline
    \\
   \multicolumn{2}{@{}l@{}}{\hbox to 0pt{\parbox{180mm}{\footnotesize
       \footnotemark[$*$] Photon index of the power-law component.
       \par\noindent
       \footnotemark[$\dagger$] Normalization in the power-law component in \\ photons keV$^{-1}$ cm$^{-2}$ s$^{-1}$ at 1 keV
     }\hss}}
      \label{tab:ps_A_pow}
    \end{tabular}
  \end{center}
\end{table}

\begin{table}[htbp]
 \caption{Flux of the point source A for each XIS sensor}
 \begin{center}
   \begin{tabular}{cc}
     \hline \hline
     sensor & flux (erg s$^{-1}$ cm$^{-2}$) \\ \hline
     XIS0   & $2.52^{+0.16}_{-0.29} \times 10^{-13}$ \\
     XIS1   & $3.19^{+0.24}_{-0.17} \times 10^{-13}$ \\
     XIS3   & $2.20^{+0.06}_{-0.15} \times 10^{-13}$ \\
     \hline
    \label{tab:ps_A_flux}
   \end{tabular}
 \end{center}
\end{table}

\begin{figure}
  \begin{center}
    \includegraphics[width=6cm, angle=270]{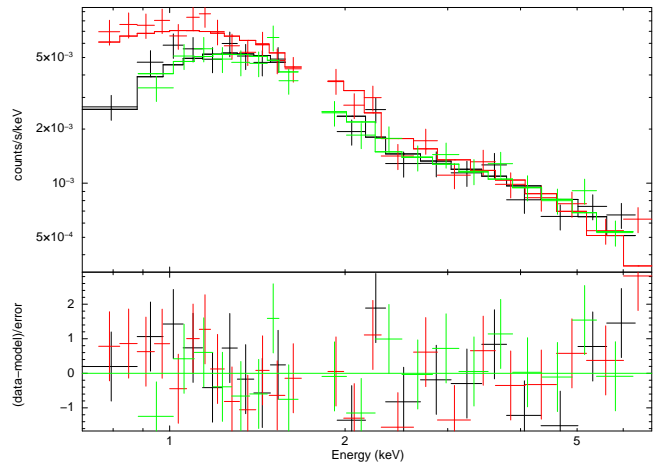} 
  \end{center}
  \caption{Same as figure \ref{fig:ps_A_pow}, but for the point source B.}
          \label{fig:ps_B_pow}
\end{figure}

The spectra of the source B are well fitted with both the powerlaw and apec model.
In the apec model fit, however, the obtained temperature is too high ($8.42^{+1.68}_{-1.08}$ keV) for galaxies.
In the powerlaw model fit, on the other hand, the resultant parameters are within a reasonable range for AGNs.
Figure \ref{fig:ps_B_pow} shows the source B spectra fitted with the power-law model and the best-fit parameters are 
listed in table \ref{tab:ps_B_pow}. The flux of the source in the energy band of 2.0-10.0 keV for each XIS sensor is 
listed in table \ref{tab:ps_B_flux}, which infers $\sim 2.5 \times 10^{-13}$ erg s$^{-1}$ cm$^{-2}$.
With NED, we check that the point source B coincides in position with 2MASX J06261933-5403158, a member galaxy of the Abell 3395. 

\begin{table}[htbp]
  \caption{Fitting results of the point source B spectra with the power-law model}
  \begin{center}
    \begin{tabular}{ccc}
      \hline \hline
      model component & parameter & value \\ \hline
      powerlaw & $\Gamma$\footnotemark[$*$] & $1.58_{-0.06}^{+0.06}$  \\
               & $N$\footnotemark[$\dagger$] & $5.81_{-0.38}^{+0.39}$ $\times$ $10^{-5}$ \\ \hline
               & $\chi^2$/d.o.f & 55.12/59   \\
      \hline
    \\
   \multicolumn{2}{@{}l@{}}{\hbox to 0pt{\parbox{180mm}{\footnotesize
       \footnotemark[$*$] Photon index of the power-law component.
       \par\noindent
       \footnotemark[$\dagger$] Normalization in the power-law component in \\ photons keV$^{-1}$ cm$^{-2}$ s$^{-1}$ at 1 keV
     }\hss}}
      \label{tab:ps_B_pow}
    \end{tabular}
  \end{center}
\end{table}

\begin{table}[htbp]
 \caption{Flux of the point source B for each XIS sensor}
 \begin{center}
   \begin{tabular}{cc}
     \hline \hline
     sensor & flux (erg s$^{-1}$ cm$^{-2}$) \\ \hline
     XIS0  & $2.48^{+0.28}_{-0.11} \times 10^{-13}$ \\
     XIS1  & $2.83^{+0.14}_{-0.11} \times 10^{-13}$ \\
     XIS3  & $2.41^{+0.14}_{-0.08} \times 10^{-13}$ \\
     \hline
    \label{tab:ps_B_flux}
   \end{tabular}
 \end{center}
\end{table}


\end{document}